\documentclass[10pt]{article}
\usepackage{amsmath}
\usepackage{amssymb}
\usepackage{amsfonts}
\usepackage{graphicx}
\usepackage{booktabs}
\usepackage{xcolor}
\usepackage{rotating}
\usepackage{setspace}
\usepackage[backend=biber, style=numeric-comp, sorting=none, autocite=superscript]{biblatex}
\usepackage{authblk}
\usepackage{pdfpages}
\usepackage[left=3cm,right=3cm,top=3cm,bottom=3cm]{geometry}

\title{Index Date Imputation for Survival Analysis in Externally Controlled Trials with Delayed Treatment Initiation}
\author[1]{Quentin Le Coent}
\author[2]{Gary L. Rosner}
\author[2]{Mei-Cheng Wang}
\author[1,2,*]{Chen Hu}
\affil[1]{Division of Quantitative Sciences, Department of Oncology, Johns Hopkins University School of Medicine, U.S.A}
\affil[2]{Department of Biostatistics, Bloomberg School of Public Health, Johns Hopkins University, U.S.A}
\affil[*]{{\em Corresponding author}: Chen Hu, huc@jhu.edu}
\date{}
\newcommand{\keywords}[1]{\textbf{Keywords:} #1}

\begin{document}
	
\maketitle
\label{firstpage}
\begin{abstract}
	Externally controlled trials compare outcomes from a single-arm trial with external controls drawn from historical trials, registries, or observational studies. For time-to-event endpoints, a key challenge arises when follow-up is indexed at treatment initiation in the single-arm trial, but the external-control data are indexed at an earlier clinical milestone, such as diagnosis or relapse. This misalignment can induce immortal time bias, distort risk sets, and complicate the interpretation of survival comparisons. We propose Index Date Imputation (IDI), a truncation-aware approach for imputing comparable index dates for external-control patients in settings with delayed treatment initiation. IDI estimates the marginal distribution of treatment-initiation times in the target single-arm population while accounting for the fact that initiation times are observed only among patients who survive long enough to initiate treatment. The imputed index dates are then used to align follow-up and enforce comparable truncation conditions in the external-control cohort. Because temporal alignment alone does not address population-level confounding, IDI is combined with propensity score weighting or matching to improve covariate comparability between cohorts. We evaluate the finite-sample performance of the proposed approach through Monte Carlo simulation studies. Using data from a randomized oncology trial, we emulate an externally controlled analysis with induced index-date misalignment and show that IDI reduces discrepancy from the randomized trial benchmark. IDI provides a practical strategy for index-date alignment in survival analyses involving delayed treatment initiation and can be integrated with standard covariate-adjustment methods when suitable external controls are available. \\
	
	\noindent \vspace{10pt}
\end{abstract}

\section{Introduction}

Externally controlled trials are increasingly used to evaluate therapeutic interventions when randomized controlled trials are infeasible, unethical, or logistically difficult, particularly in oncology and rare diseases \autocite{hatswell2016regulatory,goring2019characteristics,wang2023current}. In these studies, outcomes from a single-arm trial are compared with outcomes from external controls drawn from historical trials, registries, or real-world data. Existing methods for incorporating external controls have focused primarily on borrowing historical information or adjusting for differences in baseline prognostic factors \autocite{viele2014use,ibrahim2015power}. When the primary outcome is a time-to-event endpoint, however, ECTs pose additional methodological challenges \autocite{jahanshahi2021use}. In addition to baseline covariate comparability, validity depends on whether follow-up is indexed at comparable time origins across the single-arm and external-control cohorts.

A common challenge arises when treatment initiation in the single-arm trial occurs after a variable delay from an earlier clinical milestone, such as diagnosis, relapse, or completion of initial therapy. Patients may become eligible for the investigational regimen only after response, recovery, or survival through an initial phase of care. In contrast, external control data are often indexed at the earlier clinical milestone, where a comparable treatment-initiation time is unobserved or undefined. If treated patients must survive from the earlier milestone to treatment initiation, whereas external controls are followed from the earlier milestone, the resulting comparison includes immortal time and may distort risk sets, survival estimates, and treatment-effect estimates.

This issue has arisen in regulatory review. In the October 2022 Oncologic Drugs Advisory Committee meeting for 131I-omburtamab, the U.S. Food and Drug Administration identified index-date selection as one of several major sources of bias in an externally controlled survival comparison \autocite{fda2022briefing}. Patients in the single-arm trial were required to survive from the start of their last post-relapse treatment modality to initiation of 131I-omburtamab, whereas external controls were not required to survive an analogous interval. FDA reported that the median duration of this interval was 3.1 months; among 79 external-control patients who received any post-relapse therapy, 24 died within 3.1 months of the analogous index date. The case also illustrates that index-date alignment is only one component of a credible external-control analysis. FDA additionally raised concerns about treatment-intensity differences, calendar-time differences, small sample sizes, and unmeasured differences between the trial and external-control populations.

Existing methods for delayed treatment initiation address related but distinct problems, as discussed in Section 2. A useful distinction is between methods targeting baseline-assigned strategies and methods targeting post-initiation comparisons. We focus on the latter setting, in which follow-up is indexed at treatment initiation among patients who become eligible to receive the intervention. In externally controlled trials, this setting creates a distinct data problem: a comparable treatment-initiation time is often unavailable for external controls, and the observed initiation-time distribution in the treated cohort is distorted because initiation times are observed only among patients who survive long enough to initiate treatment. Thus, in the setting considered here, the methodological problem is not only how to align time zero, but also how to estimate the initiation-time distribution that would have been observed in the target population absent truncation.

We propose Index Date Imputation (IDI), a truncation-aware approach for imputing comparable index dates for external controls in survival analyses of externally controlled trials with delayed treatment initiation. IDI targets settings in which the scientific question is indexed at treatment initiation among patients who become eligible for a later-phase intervention, but a comparable initiation date is unavailable for external controls. Its primary methodological contribution is estimation of the marginal treatment-initiation time distribution in the target population when initiation times are observed only among treated patients who survive long enough to initiate treatment. This distinguishes IDI from resampling-based index-date emulation approaches that use the empirical distribution of observed initiation times. The imputed index dates are then used to align follow-up and apply comparable truncation conditions in the external-control cohort. Because temporal alignment alone does not address population-level confounding, IDI should be combined with covariate-adjustment methods such as propensity score weighting or matching. IDI addresses index-date alignment, a necessary but not sufficient condition for causal interpretation; valid use also requires external-control data adequate to assess eligibility at the imputed index date, sufficient covariate overlap, comparable outcome ascertainment, and adjustment for measured confounding.

The remainder of this article is organized as follows. Section 2 places the proposed approach within the target trial emulation framework and distinguishes post-initiation comparisons from baseline-assigned treatment strategies. Section 3 formalizes the externally controlled survival problem, defines the target estimand, and clarifies the role of index-date alignment. Section 4 introduces the IDI procedure and describes its integration with covariate-adjustment methods. Section 5 evaluates finite-sample performance through simulation studies. Section 6 illustrates the approach using an emulated externally controlled analysis based on a completed randomized oncology trial. Section 7 discusses practical implementation, assumptions, limitations, and implications for the design and analysis of externally controlled survival studies.
	
\section{Target Trial Emulation with Delayed Treatment Initiation} \label{sec:background}

When defining and quantifying a treatment effect using observational or otherwise non-randomized data, the target trial emulation framework provides a useful structure for specifying the comparison that an analysis seeks to approximate, including the population, eligibility criteria, treatment strategies, index date, outcome, causal contrast, and analysis plan\autocite{hernan2022target,dickerman2026target}. In settings with delayed treatment initiation, the choice of time zero is especially important because it determines the population being compared and the interpretation of the resulting survival contrast.

\begin{figure}[ht]
	\includegraphics[width=1\linewidth]{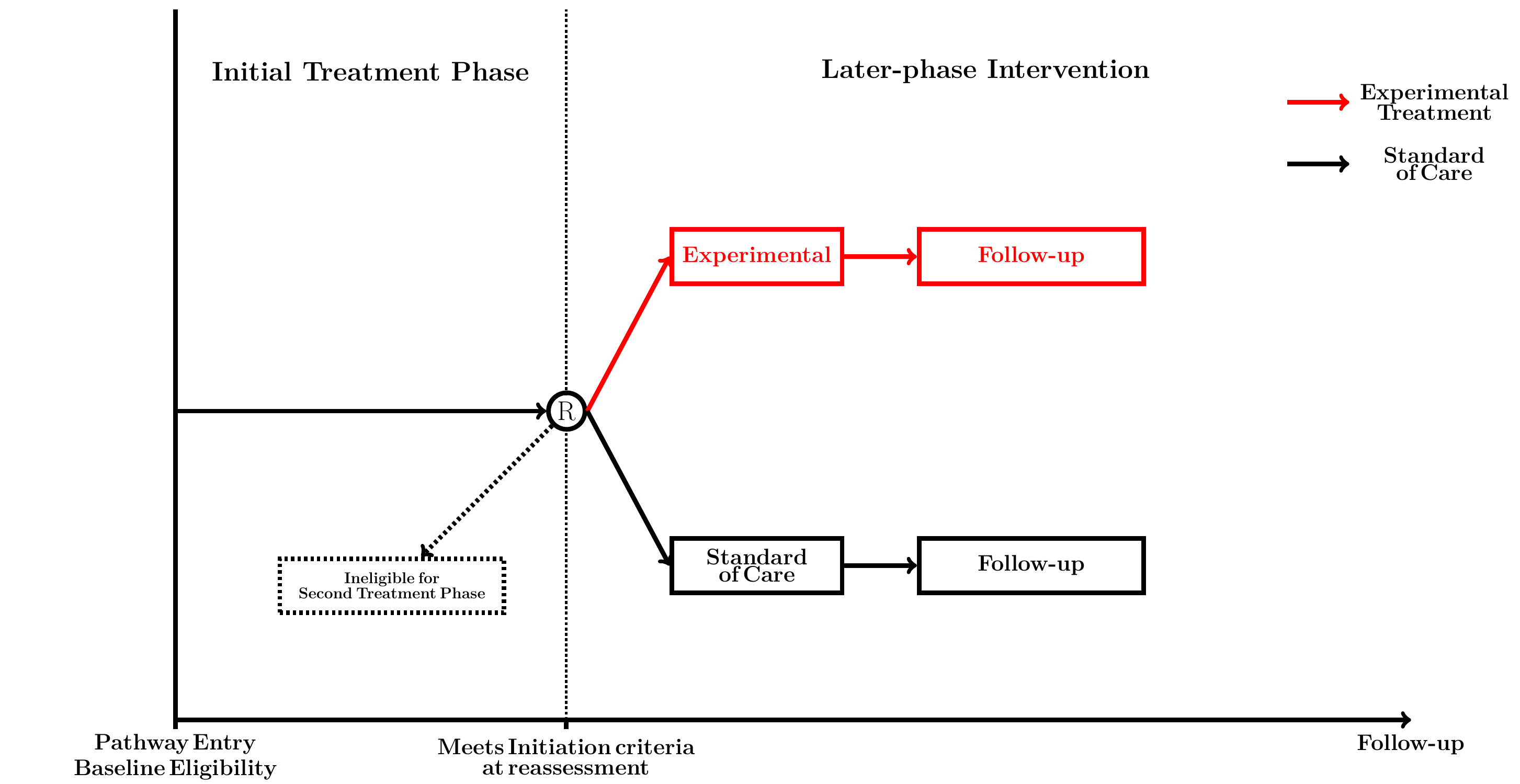}
	\caption{Target trial emulation for externally controlled trials under delayed treatment initiation. The target trial is indexed at treatment initiation among patients eligible to initiate the later-phase treatment strategy.}
	\label{fig:tte_idi}
\end{figure}

Figure~\ref{fig:tte_idi} illustrates the target randomized trial of the post-initiation comparison considered in this article. Patients first enter a common clinical pathway at an initial clinical milestone, such as diagnosis, relapse, or completion of initial therapy. During the initial phase, patients are followed while they receive the same treatment pathway or otherwise remain under the same pre-initiation clinical process. At a later decision time, denoted by \(R\), patients who remain alive and satisfy the relevant eligibility criteria become eligible to initiate the subsequent intervention. In the ideal target trial, these eligible patients would be randomized at \(R\) to initiate the investigational treatment or to receive the corresponding control strategy, and follow-up for the survival endpoint would begin at \(R\). In oncology, this formulation aligns with settings in which patients first receive induction or multimodality therapy and are then randomized, among those eligible after the initial phase, to alternative maintenance or later-phase treatment strategies. In externally controlled analyses, this formulation aligns naturally with single-arm trials in which follow-up is indexed at treatment initiation, but it requires reconstruction of a comparable decision time for external controls.

Table~\ref{tab:target_trial_idi} summarizes the main design-alignment components in this post-initiation setting. The table is intended to clarify the role of IDI within the externally controlled comparison. IDI addresses the index-date component, namely the reconstruction of a comparable decision time for external controls and the corresponding truncation condition. It does not, by itself, address all other requirements for a valid comparison. Eligibility at the imputed index date, covariate comparability, and comparable outcome ascertainment remain separate design and data requirements.

\begin{table}[t]
	\centering
	\caption{Design-alignment components for post-initiation survival comparisons with delayed treatment initiation}
	\label{tab:target_trial_idi}
	\scriptsize
	\renewcommand{\arraystretch}{1.08}
	\begin{tabular}{p{0.18\textwidth}p{0.27\textwidth}p{0.24\textwidth}p{0.22\textwidth}}
		\hline
		\textbf{Component} &
		\textbf{Target post-initiation comparison} &
		\textbf{Observed single-arm trial data} &
		\textbf{External-control implementation} \\
		\hline
		
		Population &
		Patients who enter a common pre-initiation pathway and are alive and eligible at a later decision time \(R\). &
		Patients observed to survive to treatment initiation and satisfy trial eligibility criteria at initiation. &
		External-control patients who enter an analogous pathway and can be evaluated at an imputed decision time \(R^*\). \\
		
		Index date &
		The later decision or treatment-initiation time \(R\). &
		Observed treatment-initiation date. &
		Unobserved; IDI imputes \(R^*\) using an estimated initiation-time distribution from the target single-arm population. \\
		
		Eligibility at index date &
		Eligibility for the later-phase intervention is assessed at \(R\). &
		Eligibility is known at treatment initiation by trial design. &
		Eligibility should be assessed or approximated at \(R^*\). This may require survival to \(R^*\) alone or survival plus absence of an eligibility-limiting event, such as progression before \(R^*\). \\
		
		Outcome follow-up &
		Time from \(R\) to the event of interest, censoring, or administrative end of follow-up. &
		Time from observed treatment initiation to event or censoring. &
		Time from \(R^*\) to event or censoring among patients satisfying the corresponding truncation and eligibility conditions. \\
		
		Covariate comparability &
		Baseline and pre-initiation prognostic factors are comparable across groups. &
		Covariates are measured in the single-arm trial cohort. &
		Covariate imbalance is addressed using weighting, matching, or related adjustment methods. \\
		
		Role of IDI &
		A comparable index date is required before post-initiation survival can be compared across groups. &
		Observed initiation times are available only among patients who survive long enough to initiate treatment. &
		IDI estimates the marginal initiation-time distribution while accounting for truncation in the observed single-arm initiation times, then uses this distribution to impute \(R^*\). \\
		
		\hline
	\end{tabular}
\end{table}

Existing methods for delayed treatment initiation address related but distinct problems. Landmark analysis, left truncation, and time-dependent regression approaches can reduce certain forms of time-related bias, but they do not generally reconstruct a comparable treatment-initiation time for external controls \autocite{anderson1983analysis,morgan2019landmark,wang1993statistical,fisher1999time}. Clone-censor-weighting and sequential trial emulation are useful when the scientific question concerns strategies assigned at an initial baseline time, for example strategies that involve initiating treatment within a prespecified window \autocite{maringe2020reflection,gaber2024mystifying,hernan2016using}. In contrast, the setting considered here is indexed at the later decision time \(R\), after patients have become eligible for the later-phase intervention. Thus, the primary problem is not how to compare baseline-assigned strategies, but how to reconstruct an analogous post-eligibility index date for external controls.

A closer class of approaches includes prescription-time distribution matching (PTDM), originally developed in pharmacoepidemiology to address immortal time bias by aligning treatment initiation times across exposure groups \autocite{zhou2005survival}. In PTDM, pseudo-initiation times are assigned to control patients by sampling from the empirical distribution of initiation times observed in the treated cohort. Related index-date emulation and multiple-imputation approaches have recently been considered in externally controlled studies and precision medicine evaluations \autocite{antunes2026choosing,weymann2025addressing}. These approaches aim to approximate the post-initiation risk sets observed in the treated cohort by assigning an analogous index date to control patients.

The key distinction motivating IDI is that, in the present setting, observed treatment-initiation times in the single-arm trial are themselves subject to truncation. A patient contributes an observed initiation time only if the patient survives and remains eligible long enough to initiate treatment. Consequently, the empirical distribution of observed initiation times estimates a survivor-selected distribution rather than the marginal initiation-time distribution in the target trial population. This distinction is small when few patients are truncated, but it can become substantial when many patients die, progress, or are otherwise censored before treatment initiation. Section~C of the Supporting Information provides a focused comparison between the proposed method and PTDM-type imputation across different degrees of truncation.

These considerations motivate a truncation-aware approach to index-date alignment. In the externally controlled setting considered here, the scientific question concerns post-initiation survival among patients who become eligible to receive the investigational regimen after satisfying a post-baseline criterion. The comparable decision time \(R\) is observed in the single-arm trial but is unobserved or undefined in the external-control cohort, and the observed distribution of \(R\) in the single-arm trial is distorted by survival and eligibility to initiation. IDI addresses this specific data problem by estimating the initiation-time distribution while accounting for truncation and using this distribution to impute comparable index dates for external controls.

\section{Formal Problem Setup and Target Estimand} 
\label{sec:objective}

\begin{figure}
	\centering
	\includegraphics[width=0.7\linewidth]{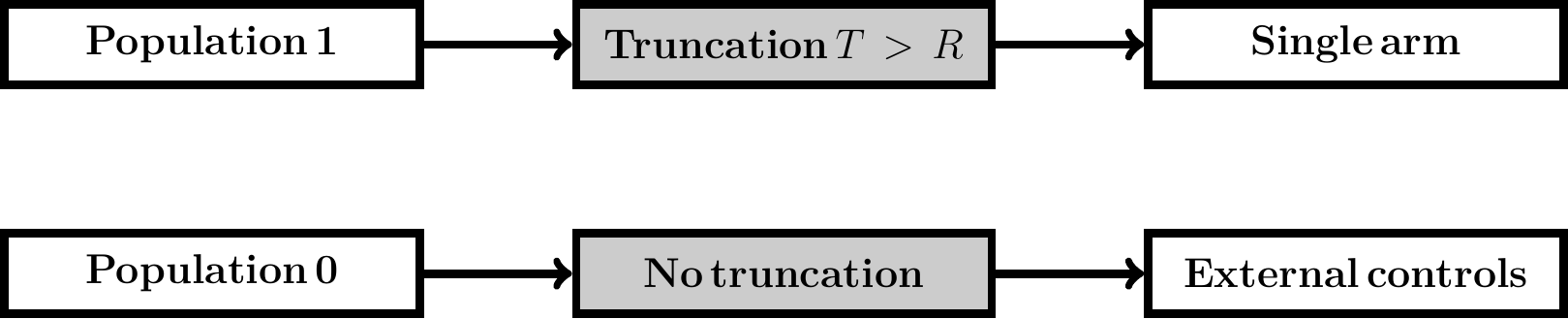}\vspace{0.5cm}
	\includegraphics[width=0.8\linewidth]{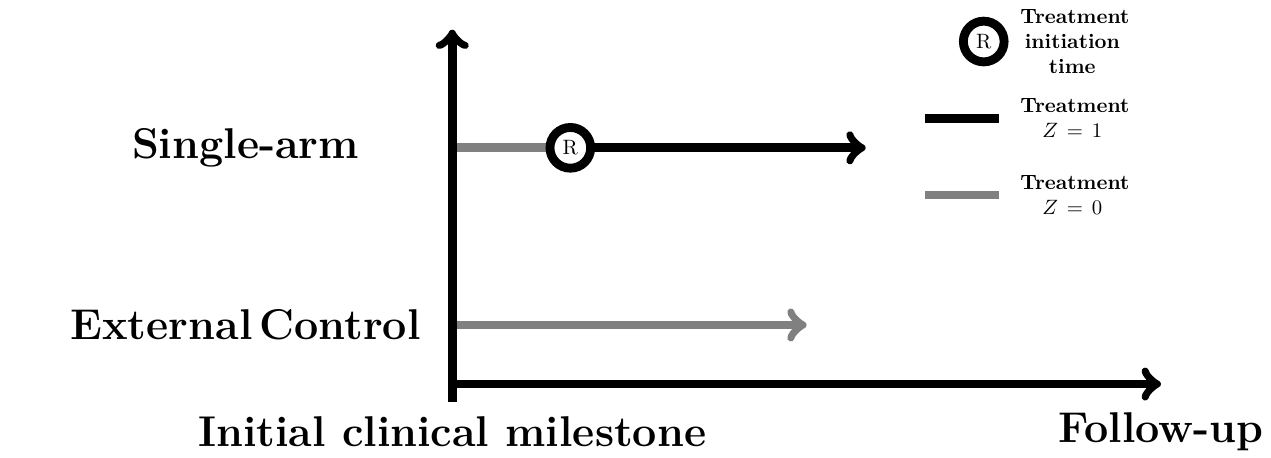}
	\caption{Schematic of data structure in an externally controlled comparison. Top: sampling mechanism. Bottom: observed variables. \label{fig:case_generic}}
\end{figure}

We consider an externally controlled setting in which patients enrolled in a single-arm trial are compared with an external-control cohort of patients who do not receive the experimental treatment. The general setting is illustrated in Figure~\ref{fig:case_generic}. Patients in both cohorts enter a clinical pathway at an initial milestone, such as diagnosis, relapse, or completion of initial therapy. In the single-arm trial, patients initiate the experimental treatment only after a delay from this initial milestone, and survival is anchored at the treatment-initiation time. In the external-control cohort, follow-up is commonly indexed at the initial milestone, and a comparable treatment-initiation time is unobserved or undefined. Moreover, patients observed in the single-arm trial are those who survive and remain eligible long enough to initiate treatment, creating a truncation mechanism that must be accounted for when reconstructing comparable risk sets.

Let \(R\) denote the post-baseline decision time at which a patient would become eligible to initiate the later-phase intervention in the target comparison, measured from the initial clinical milestone. In the observed single-arm trial, \(R\) is observed as the treatment-initiation date. In the external-control cohort, the analogous \(R\) is unobserved and must be reconstructed. We assume that \(R\) is determined before assignment to the later-phase treatment strategy, under the common pre-initiation clinical pathway. Accordingly, \(R\) is not indexed by treatment strategy. This convention reflects the post-initiation target trial in Figure~\ref{fig:tte_idi}, where treatment strategies diverge at \(R\).

Let \(E(R)\) denote the indicator that a patient is alive and satisfies the relevant later-phase eligibility criteria at decision time \(R\). In the simplest setting, eligibility at \(R\) requires only survival to \(R\), so that \(E(R)=\mathbb{I}\{T>R\}\). In other settings, eligibility may also depend on additional clinical events observed before \(R\). For example, in a maintenance-therapy setting, progression before \(R\) may preclude later-phase eligibility; if \(P\) denotes time to progression from the initial milestone, the relevant eligibility condition may be written as \(E(R)=\mathbb{I}\{T>R, P>R\}\). This notation separates the decision time \(R\) from the clinical information needed to determine whether a patient belongs to the post-initiation risk set at that time.

Let \(n_1\) and \(n_0\) denote the sample sizes in the single-arm trial and external-control cohorts, respectively. Let \(G \in \{0,1\}\) denote cohort membership, with \(G=1\) for the single-arm trial population and \(G=0\) for the external-control population. Let \(Z \in \{0,1\}\) denote the treatment strategy, with \(Z=1\) for the experimental treatment and \(Z=0\) for the control strategy. In the externally controlled setting considered here, all patients with \(G=1\) receive \(Z=1\), and all patients with \(G=0\) receive \(Z=0\). We distinguish \(G\) from \(Z\) because \(G\) indexes cohort membership and sampling, whereas \(Z\) indexes the treatment strategy being compared.

Let \(X\) denote a vector of covariates measured at the initial pathway-entry time. Let \(T(z)\) denote the event time, measured from pathway entry, under treatment strategy \(z \in \{0,1\}\), and let \(C\) denote the censoring time. Because the treatment strategies differ only from the post-baseline decision time \(R\) onward, the pre-initiation pathway is common across treatment strategies up to \(R\). Under consistency, the observed outcome in the single-arm trial is \(Y=\min\{T(1),C\}\), with event indicator \(\delta=\mathbb{I}\{T(1)\leq C\}\), and the observed data include \((X,Y,\delta,R,G,Z)\). In the external-control cohort, the observed outcome is \(Y=\min\{T(0),C\}\), with event indicator \(\delta=\mathbb{I}\{T(0)\leq C\}\), and the observed data include \((X,Y,\delta,G,Z)\), with \(R\) unobserved.

The target comparison underlying the proposed framework is indexed at the post-baseline decision time \(R\), rather than at pathway entry. The target population is the population represented by the single-arm trial that would remain alive and satisfy the relevant eligibility criteria at \(R\), denoted by \(E(R)=1\). The treatment strategies are initiation of the experimental treatment at \(R\) versus receipt of the corresponding control strategy from \(R\) onward. The endpoint is time from \(R\) to the event of interest. The primary survival functions of interest are the marginal post-initiation survival functions in the target single-arm population,
\begin{equation}
	\label{eq:msurv}
	S_z(t) \equiv \mathbb{P}\{T(z)-R > t \mid E(R)=1, G=1\}, 
	\quad z\in \{0,1\}.
\end{equation}
Here, \(S_1(t)\) denotes the survival function that would be observed from \(R\) onward under initiation of the experimental treatment, and \(S_0(t)\) denotes the corresponding survival function under the control strategy, both in the \(G=1\) target population. The conditioning event \(E(R)=1\) reflects membership in the post-initiation target population. It is part of the comparison being emulated, rather than a post hoc analytic exclusion. In applications where eligibility at \(R\) depends on clinical criteria beyond survival, those criteria must be emulated or approximated in the external-control cohort when such data are available.

The survival function \(S_1(t)\) is estimable from the observed single-arm trial data using standard survival methods, provided censoring is appropriately handled. In contrast, \(S_0(t)\) is not directly estimable from the external-control cohort without additional adjustment. The external-control cohort is drawn from \(G=0\), rather than the target population \(G=1\); its follow-up is indexed at pathway entry rather than at \(R\); and the comparable post-initiation risk-set condition \(E(R)=1\) cannot be assessed unless an analogous \(R\) is reconstructed. Without adjustment, the external-control data identify survival under the control strategy from pathway entry in the external-control population, rather than the post-initiation survival function in Equation~\ref{eq:msurv}.

Recovering \(S_0(t)\) from the external-control cohort therefore requires three components. First, the covariate distribution of the external-control cohort must be aligned with that of the target single-arm population. Second, a comparable decision time \(R\) must be assigned to external-control patients so that follow-up can be re-indexed from pathway entry to \(R\). Third, external-control patients must satisfy the corresponding risk-set and eligibility condition \(E(R)=1\), including survival to \(R\) and, when relevant data are available, absence of eligibility-limiting events before \(R\). The method developed below uses truncation-aware estimation of the initiation-time distribution to support index-date imputation and risk-set alignment, and combines this step with propensity score weighting or matching to address measured population-level differences between \(G=0\) and \(G=1\).

\section{Methods} 
\label{sec:methods}

We now describe the Index Date Imputation (IDI) procedure for estimating the post-initiation control survival function \(S_0(t)\) in Equation~\ref{eq:msurv} using the external-control cohort. Among the design-alignment components summarized in Table~\ref{tab:target_trial_idi}, IDI addresses the index-date component and the associated post-initiation risk-set condition. The remaining components, including assessment of eligibility at the imputed index date, covariate comparability, and comparable outcome ascertainment, are required for implementation but are not themselves solved by index-date imputation.

The procedure has two distinct roles. First, it reconstructs comparable index dates for external-control patients and enforces the corresponding post-initiation risk-set and eligibility condition. Second, it combines this temporal alignment step with covariate adjustment to target the \(G=1\) population. Throughout, IDI refers to the full analytic procedure, with index-date imputation serving as the central component. The procedure consists of four steps: (i) estimation of the initiation-time distribution and propensity score weights, (ii) imputation of pseudo-initiation times for external controls, (iii) estimation of marginal survival contrasts, and (iv) bootstrap inference.

The procedure relies on standard design and identification assumptions. First, after adjustment for measured covariates \(X\), the external-control cohort contains sufficient information to represent the control survival experience that would have been observed in the \(G=1\) target population. Second, there is adequate covariate overlap between the single-arm and external-control cohorts. Third, censoring is independent of the event time conditional on variables used in the analysis. Fourth, the post-baseline decision time \(R\) is determined before assignment to the later-phase treatment strategy, and treatment strategies differ only after \(R\). Fifth, external-control patients retained after imputation should satisfy the same risk-set and eligibility condition at the imputed decision time as patients in the target post-initiation comparison. In the notation of Section~\ref{sec:objective}, this condition is denoted by \(E(R)=1\).

For the nonparametric estimator of the marginal initiation-time distribution used below, we additionally impose the working assumption that \(R\) is independent of \(X\). This assumption is not part of the definition of the post-initiation estimand, but is used to estimate the marginal distribution \(F_R\) from truncated single-arm data. If initiation timing depends on covariates, a conditional model for \(R \mid X\) that accounts for truncation would be required. We evaluate the performance of the proposed procedure under both covariate-independent and covariate-dependent initiation times in Section~\ref{sec:simulations}.

\subsection{Step 1: Estimation of the Initiation-Time Distribution and Propensity Score Weights}

To target the post-initiation average treatment effect on the treated (ATT) in the \(G=1\) population, we estimate a propensity score for cohort membership,
\[
p(X)=\mathbb{P}(G=1\mid X),
\]
where \(G=1\) denotes membership in the single-arm trial population. This is a data-source or population-membership propensity score rather than a treatment-assignment propensity score. In the single-arm ECT setting considered here, \(G\) and \(Z\) are perfectly aligned in the observed data, but \(p(X)\) is used to balance measured covariates between the external-control cohort and the target single-arm population. We estimate \(p(X)\) using a logistic regression model,
\begin{equation}
	\label{eq:ps_model}
	\mathbb{P}(G=1 \mid X) =
	\frac{\exp(\beta_0+\beta^\top X)}
	{1+\exp(\beta_0+\beta^\top X)}.
\end{equation}

A complication is that the observed single-arm cohort is a truncated sample of the target \(G=1\) population: initiation times are observed only for patients who survive and remain eligible long enough to initiate treatment. A naive propensity score model that pools the observed single-arm patients with external controls would therefore estimate covariate balance for the survivor-selected single-arm cohort, rather than for the target population at pathway entry. To account for this selection, single-arm patients are weighted by the inverse of their probability of satisfying the later-phase risk-set and eligibility condition at the observed initiation time.

Let,
\[
Q_E(r)=\mathbb{P}\{E(r)=1\mid G=1\},
\]
denote the marginal probability that a patient in the target single-arm population remains alive and eligible at time \(r\) under the common pre-initiation pathway. In the survival-only setting, \(E(r)=\mathbb{I}\{T(1)>r\}\), so that \(Q_E(r)=S_{T(1)}(r\mid G=1)\). When additional eligibility-limiting events are observed, \(Q_E(r)\) can be defined accordingly. For example, if progression before maintenance therapy precludes later-phase eligibility and \(P\) denotes time to progression, then \(E(r)=\mathbb{I}\{T(1)>r,P>r\}\) and \(Q_E(r)=\mathbb{P}\{T(1)>r,P>r\mid G=1\}\).

Under the working assumption that \(R\) is independent of \(X\), the marginal distribution \(F_R\) can be estimated nonparametrically from the observed single-arm initiation times after correcting for truncation by \(E(R)=1\). We use the weighted empirical estimator
\begin{equation}
	\label{eq:ghat}
	\hat F_R(r) =
	\frac{\sum_{i:G_i=1} \hat Q_E^{-1}(R_i)\,\mathbb{I}(R_i \le r)}
	{\sum_{i:G_i=1} \hat Q_E^{-1}(R_i)}.
\end{equation}
In the survival-only case, \(\hat Q_E(r)\) is the left-truncated Kaplan-Meier estimator \(\hat S_{T(1)}(r\mid G=1)\), and Equation~\eqref{eq:ghat} reduces to the inverse-survival weighted estimator
\[
\hat F_R(r) =
\frac{\sum_{i:G_i=1} \hat S_{T(1)}^{-1}(R_i \mid G=1)\,\mathbb{I}(R_i \le r)}
{\sum_{i:G_i=1} \hat S_{T(1)}^{-1}(R_i \mid G=1)}.
\]
This estimator corresponds to the nonparametric maximum likelihood estimator of \(F_R\) under the independent truncation model \autocite{wang1991nonparametric}. Section~C of the Supporting Information contrasts this estimator with the empirical distribution of observed initiation times used in PTDM-type approaches, and illustrates how the difference between the truncated and untruncated initiation-time distributions increases as truncation becomes more severe.

The same eligibility notation is also useful for constructing the inverse truncation weights used in the propensity score model. For a single-arm patient with covariates \(X_i\), the probability of being observed in the post-initiation single-arm cohort is
\begin{equation}
	\label{eq:truncprob_general}
	\pi_E(X_i)
	=
	\mathbb{P}\{E(R)=1\mid X_i\}
	=
	\int \mathbb{P}\{E(r)=1\mid X_i\}\,\mathrm{d}F_R(r).
\end{equation}
In the survival-only setting, this becomes
\[
\mathbb{P}\{T(1)>R\mid X_i\}
=
\int \mathbb{P}\{T(1)>r\mid X_i\}\,\mathrm{d}F_R(r).
\]
We estimate \(\mathbb{P}\{T(1)>r\mid X\}\) using a working Cox proportional hazards model with a Breslow baseline hazard estimator:
\[
\hat{\mathbb{P}}\{T(1)>r \mid X\}
=
\exp\!\left\{-\hat{\Lambda}_0(r)\exp(\hat{\theta}^\top X)\right\},
\]
where
\[
\hat{\Lambda}_0(r)
=
\sum_{T_i(1)\le r}
\frac{\delta_i}
{\sum_{j\in\mathcal{R}\{T_i(1)\}}\exp(\hat{\theta}^\top X_j)},
\]
and \(\mathcal{R}(t)\) denotes the risk set at time \(t\). Integrating over \(\hat F_R\) yields
\[
\hat \pi_E(X_i)
=
\int \hat{\mathbb{P}}\{T(1)>r\mid X_i\}\,\mathrm{d}\hat F_R(r)
\]
in the survival-only case, and the corresponding inverse truncation weight is
\[
\zeta_i = \hat\pi_E(X_i)^{-1}.
\]
When additional eligibility-limiting events are included in \(E(r)\), the same expression is used with \(\hat{\mathbb{P}}\{E(r)=1\mid X_i\}\) replacing \(\hat{\mathbb{P}}\{T(1)>r\mid X_i\}\). This extension changes only the truncation probability used for weighting and estimation of \(F_R\); the role of IDI remains the same.

Intuitively, patients with a lower probability of surviving and remaining eligible to treatment initiation are underrepresented in the observed single-arm cohort and therefore receive larger weights when reconstructing the covariate distribution of the target \(G=1\) population at pathway entry. These inverse truncation weights are used for estimating the population-membership propensity score, not for redefining the observed post-initiation survival experience of the treated cohort in the primary ATT analysis.

The inverse truncation weights \(\zeta_i\) are applied to single-arm patients when fitting the propensity score model in Equation~\eqref{eq:ps_model}. External-control patients receive unit weights in this propensity score model because they are sampled at pathway entry rather than conditional on survival and eligibility to experimental treatment initiation. After estimating \(\hat p_i=\widehat{\mathbb{P}}(G_i=1\mid X_i)\), the ATT weights are
\begin{equation}
	\label{eq:attweight}
	w_i =
	\begin{cases}
		1, & G_i = 1, \\[6pt]
		\dfrac{\hat p_i}{1-\hat p_i}, & G_i = 0 .
	\end{cases}
\end{equation}

\subsection{Step 2: Imputation of Pseudo-Initiation Times}

To align time zero in the external-control cohort, we impute pseudo-initiation times from the estimated initiation-time distribution \(\hat F_R\). For each external-control patient \(j=1,\ldots,n_0\), we draw \(\hat R_j \sim \hat F_R\), compute the re-indexed follow-up time \(\tilde Y_j=Y_j-\hat R_j\), and retain the patient only if the post-initiation risk-set and eligibility condition is satisfied at \(\hat R_j\).

In the survival-only setting, this retention step is implemented by requiring \(Y_j>\hat R_j\), which enforces observed follow-up beyond the imputed index date. Patients whose follow-up ends before the imputed index date are not included in the post-initiation risk set. When trial eligibility at \(R\) depends on clinical criteria beyond survival, those criteria should also be applied to the external-control cohort at the imputed index date whenever the required data are available. For example, if trial eligibility requires absence of disease progression before treatment initiation, then external-control patients who progressed before the imputed index date should be excluded. If these criteria cannot be evaluated in the external-control data, IDI can still align time zero but does not fully reproduce the post-initiation eligibility process. The analysis should then be interpreted as an approximation to the target comparison rather than a complete emulation.

\subsection{Step 3: Estimating Marginal Survival Contrasts}

After Steps 1 and 2, the adjusted external-control cohort is restricted to patients satisfying the post-initiation risk-set and eligibility condition and re-indexed at the imputed decision time \(\hat R\). Because ATT weights are used to make the external-control cohort represent the covariate distribution of the \(G=1\) target population, \(S_0(t)\) should be estimated using a weighted Kaplan-Meier estimator or another weighted survival estimator. The estimator is applied to \(\tilde Y_i=Y_i-\hat R_i\) among retained external-control patients, with the original event indicator \(\delta_i\) and weights \(w_i\). The experimental survival function \(S_1(t)\) is estimated from the single-arm trial using \(\tilde Y_i=Y_i-R_i\), with unit weights in the primary ATT analysis.

The inverse truncation weights from Step~1 are used to estimate the population-membership propensity score after accounting for survival-based or eligibility-based sampling into the single-arm cohort. They are not used as outcome-analysis weights for \(S_1(t)\) in the primary ATT analysis, because the observed single-arm patients define the treated post-initiation target population.

Treatment effects may be summarized using contrasts of the estimated marginal survival functions, such as survival probability differences at fixed times or restricted mean survival time differences. A weighted Cox proportional hazards model may also be used as a model-based summary,
\[
\lambda(t \mid Z)=\lambda_0(t)\exp(\alpha \cdot Z),
\]
where \(t\) is measured from the observed or imputed index date. The resulting coefficient \(\hat\alpha\) should be interpreted as a weighted, model-based log hazard ratio comparing the aligned marginal survival curves. Because hazard ratios are non-collapsible and require additional modeling assumptions, survival probability or restricted mean survival time contrasts may provide more direct marginal summaries.

\subsection{Step 4: Nonparametric Bootstrap for Inference}

The IDI procedure includes stochastic imputation of index dates and estimation of nuisance quantities, including \(\hat F_R\), truncation probabilities, and propensity score weights. To account for these sources of variability, we use a nonparametric bootstrap \autocite{mooney1993bootstrapping}. For \(b=1,\ldots,B\), we resample the single-arm and external-control cohorts with replacement, re-estimate all nuisance quantities, re-impute pseudo-initiation times for external controls, reconstruct the adjusted analysis cohort, and recompute the treatment-effect summary of interest. The empirical distribution of the resulting estimates is used to obtain standard errors and confidence intervals, such as percentile intervals.

Diagnostic tools for assessing implementation of the IDI procedure are described in Section~A of the Supporting Information. These include standardized mean differences for evaluating covariate balance before and after propensity score adjustment, and a Q-Q plot comparing the estimated truncated version of \(F_R\) with the empirical distribution of observed initiation times in the single-arm cohort. A summary flowchart of the IDI procedure is provided in Figure~3 of the Supporting Information.

IDI can be combined with either propensity score weighting or propensity score matching to address measured baseline differences between \(G=0\) and \(G=1\). The weighting implementation uses ATT weights as described above. Alternatively, 1:1 nearest-neighbor matching can be used to construct a matched subset of external controls, after which the index-date imputation and risk-set alignment steps are applied as described above, with unit weights used in the outcome model. In the simulation study, we evaluate both implementations.

\section{Monte Carlo Simulations} 
\label{sec:simulations}

We conducted Monte Carlo simulations to evaluate the operating characteristics of the proposed IDI procedure in externally controlled single-arm trials with delayed treatment initiation. The simulations had two objectives. The primary simulation evaluated whether index date imputation, when combined with propensity score matching or weighting, reduces bias from index-date misalignment, survival-based truncation, and baseline covariate imbalance. We also included component analyses to separate the contributions of temporal alignment and covariate adjustment. A focused secondary simulation compared IDI with PTDM-type index-date imputation under increasing degrees of truncation, isolating the setting in which the empirical distribution of observed initiation times differs from the marginal initiation-time distribution.

For the primary simulation, we evaluated performance under two initiation-time mechanisms: one in which the initiation time \(R\) is independent of baseline covariates, consistent with the working assumption used for the nonparametric estimator of \(F_R\), and one in which \(R\) depends on baseline covariates, representing a robustness scenario in which this working assumption is violated. The overall data-generating mechanism is illustrated in Figure~4 of the Supporting Information.

\subsection{Data Generation}

We simulated an underlying population from which single-arm trial and external-control cohorts were sampled. Each subject had baseline covariates \(X_i=(X_{i1},X_{i2})\), where \(X_{i1}\sim\mathrm{Bernoulli}(0.5)\) and \(X_{i2}\sim\mathcal{N}(0,0.3^2)\). Cohort membership was generated according to
\[
\operatorname{logit}\!\left\{\mathbb{P}(G_i=1\mid X_i)\right\}
=
\beta_0+\beta_1X_{i1}+\beta_2X_{i2},
\]
with \(\beta_0=-0.5\), \(\beta_1=0.3\), and \(\beta_2=-0.2\). From this generated population, we sampled \(n_1=200\) patients for the single-arm trial cohort \((G=1)\) and \(n_0=800\) patients for the external-control cohort \((G=0)\). The treatment indicator \(Z_i\) was set to 1 for patients in the single-arm trial and 0 for patients in the external-control cohort.

For patients in the single-arm trial population, we considered two mechanisms for the treatment-initiation time \(R_i\):
\begin{enumerate}
	\item \textbf{Covariate-independent initiation:} \(R_i\sim\mathcal{U}(0,2)\);
	\item \textbf{Covariate-dependent initiation:} \(R_i\) was generated from an exponential distribution with hazard
	\[
	\lambda_R(r\mid X_i)
	=
	\lambda_{0,R}\exp(\beta_{R,1}X_{i1}+\beta_{R,2}X_{i2}),
	\]
	where \(\lambda_{0,R}=0.5\), \(\beta_{R,1}=-2\), and \(\beta_{R,2}=-2\).
\end{enumerate}
The second scenario was included to assess robustness when initiation timing depends on baseline covariates, a setting in which the nonparametric estimator of the marginal initiation-time distribution is misspecified unless a conditional model for \(R\mid X\) is used.

For patients in the single-arm population, survival time from pathway entry, \(T_i(1)\), was generated from a piecewise exponential model with hazard
\[
\lambda_{T(1)}(t\mid X_i,R_i)
=
\lambda_0\exp\!\left\{\gamma^\top X_i+\alpha\,\mathbb{I}(t\ge R_i)\right\},
\]
so that the treatment effect began at the initiation time \(R_i\). Patients with \(T_i(1)\le R_i\) were excluded from the observed single-arm cohort, inducing survival-based truncation. Censoring was generated as \(C_i=R_i+W_i\), where \(W_i\sim\mathrm{Exp}(0.3)\). For external controls, treatment-initiation times were unobserved and survival times were generated from
\[
\lambda_{T(0)}(t\mid X_i)
=
\lambda_0\exp(\gamma^\top X_i),
\]
with censoring \(C_i\sim\mathrm{Exp}(0.3)\). The same baseline hazard \(\lambda_0\) was used for both groups. Observed follow-up times were \(Y_i=\min\{T_i(1),C_i\}\), with event indicator \(\delta_i=\mathbb{I}\{T_i(1)\le C_i\}\), for single-arm patients, and \(Y_i=\min\{T_i(0),C_i\}\), with event indicator \(\delta_i=\mathbb{I}\{T_i(0)\le C_i\}\), for external controls.

In the observed data, external controls had only follow-up time indexed at pathway entry, \(Y_i\). In contrast, single-arm trial patients had both the pathway-entry-indexed follow-up time \(Y_i\) and the observed treatment-initiation time \(R_i\), allowing follow-up to be re-indexed as \(Y_i-R_i\). The average number of observed events in each group before adjustment is reported in Table~4 of the Supporting Information. Figure~5 of the Supporting Information illustrates the conditional hazard functions and marginal survival functions under the null treatment effect \((\alpha=0)\), before adjustment for truncation or confounding.

For the primary simulation, we compared five analysis strategies. The first was a naive Cox model indexed at pathway entry, with no covariate adjustment and no index-date alignment. This analysis is expected to be biased because single-arm patients must survive to treatment initiation, whereas external controls are analyzed from pathway entry. The second strategy used propensity score weighting without imputing index date, retaining the pathway-entry time origin. This strategy evaluates whether covariate adjustment alone can address the bias induced by delayed treatment initiation. The third strategy used index date imputation without propensity score adjustment, thereby evaluating the contribution of index-date alignment alone. The fourth strategy used propensity score matching followed by index date imputation. The fifth strategy used propensity score weighting followed by index date imputation. In all index date imputation-based strategies, pseudo-initiation times were imputed for retained external-control patients, follow-up was re-indexed at the observed or imputed initiation time, and the post-initiation comparison was summarized using a Cox model. Although survival was generated from a piecewise exponential model with a hazard change at \(R_i\), the Cox model was used as a model-based summary of the aligned marginal survival contrast, not because proportional hazards were assumed to hold exactly.

For each scenario, the true value reported in the simulation tables corresponds to the model-based marginal log-hazard ratio obtained under the same post-initiation target contrast in a large simulated population. We evaluated performance using the mean estimate, absolute bias, empirical standard deviation, average estimated standard error, and empirical coverage of the nominal 95\% confidence interval. Each scenario was evaluated over \(10^4\) Monte Carlo replicates under three treatment-effect settings: \(\alpha=0\), \(\alpha=-0.5\), and \(\alpha=-1\).

\subsection{Simulation Results}

Results from the primary simulation with \(n_1=200\) and \(n_0=800\) are presented in Table~\ref{tab:res_ptdm_s1}. Results for smaller sample sizes \((n_1=100,n_0=400)\) are reported in Table~5 of the Supporting Information. Across all scenarios, the naive baseline-indexed Cox model showed substantial bias and poor coverage, reflecting the combined effects of index-date misalignment, survival-based truncation, and baseline covariate imbalance.

The component analyses illustrate the distinct roles of temporal alignment and covariate adjustment. Propensity score weighting without index date imputation performed equally poorly compared to the naive analysis. As a result, this strategy is expected to retain substantial bias when the dominant problem is index-date misalignment. Index date imputation without propensity score adjustment corrected the index-date alignment problem but did not fully address differences between the single-arm and external-control covariate distributions. 
In the covariate-independent truncation scenario, its performance was comparable to the combined approaches, but in the covariate-dependent truncation scenario this approach resulted in a larger bias and a reduced coverage rate of the confidence interval. 
The combined approaches, particularly PS weighting plus index date imputation, addressed both sources of bias and produced the most stable performance across the main scenarios.

Accross all scenarios, propensity score weighting combined with index date imputation had smaller empirical standard deviation than propensity score matching combined with index date imputation, while maintaining minimal bias and approximately nominal coverage. Under covariate-dependent initiation, both index date imputation-based approaches remained approximately unbiased in the scenarios considered, although variability was higher than in the covariate-independent setting. This scenario should be interpreted as a robustness assessment because the nonparametric estimator of \(F_R\) used in IDI assumes initiation times are independent of baseline covariates. 

Estimated power curves based on the model-based Cox summary are shown in Figure~6 of the Supporting Information. The top panel corresponds to covariate-independent initiation, and the bottom panel corresponds to covariate-dependent initiation. Across the simulated settings, index date imputation combined with propensity score weighting achieved higher power than index date imputation combined with matching, reflecting the larger effective sample size retained under weighting compared to the 1:1 matching.

We also conducted a focused comparison between IDI and PTDM-type index-date imputation, with detailed results provided in Section~C of the Supporting Information. This comparison was designed to isolate the index-date imputation component and therefore did not include covariates, censoring, or a treatment effect. In this simplified setting, any bias arises from the distribution used to impute external-control index dates rather than from confounding, censoring, or model misspecification. The single-arm cohort was sampled conditional on surviving beyond treatment initiation, whereas the external-control cohort was sampled without this truncation. PTDM-type imputation assigned pseudo-index dates from the empirical distribution of observed initiation times in the single-arm cohort, while IDI assigned pseudo-index dates from the estimated marginal initiation-time distribution after correcting for truncation.

The focused comparison showed that IDI and PTDM-type imputation performed similarly when truncation was mild. However, as truncation became more severe, the empirical distribution of observed initiation times increasingly differed from the marginal initiation-time distribution. PTDM-type imputation then generated systematically earlier pseudo-index dates for external controls, which artificially lengthened external-control survival measured from the imputed index date and produced increasing bias. In contrast, IDI remained approximately unbiased across the truncation scenarios. These results support the central role of truncation-aware estimation of \(F_R\) when observed initiation times are materially survivor-selected.

Sensitivity analyses assessing robustness to unmeasured confounding are provided in Section~B of the Supporting Information. As expected, stronger unmeasured confounding resulted in large bias and poor coverage rate of the confidence intervals for index date imputation combined with either propensity score weighting or matching., reinforcing that IDI, like other propensity score based approaches, depends on adequate adjustment for prognostic factors that differ between the single-arm and external-control cohorts.

The simulations focus on the core methodological problem of truncation-aware index-date alignment and its combination with measured covariate adjustment. Other implementation issues in external-control studies, including limited covariate overlap, incomplete longitudinal eligibility information, and eligibility misclassification at the imputed index date, require study-specific diagnostics and sensitivity analyses and are discussed as practical limitations rather than treated as separate simulation factors here.

\begin{table}
\caption{Simulation Results for IDI ($n_1 = 200$, $n_0 = 800$) \label{tab:res_ptdm_s1}}
\centering
\scalebox{1}{
	\begin{tabular}{lcccccc}
		\hline  \addlinespace[0.3em]
		Method & True & Mean$^{*}$ & Abs. Bias$^{**}$ & SD$^{***}$ & SE$^{\dagger}$ & Cov.$^{\ddagger}$ \\ 
		\addlinespace[0.3em]
		\hline \hline \addlinespace[0.3em]
		
		\multicolumn{7}{c}{\textbf{First Case: Covariate-independent truncation} ($R \perp X$)} \\
		\addlinespace[0.3em] \hline \hline \addlinespace[0.3em]
		
		\multicolumn{7}{c}{Treatment effect $= 0$} \\ \addlinespace[0.3em]
		
		Naive analysis & 0.000 & -0.700 & 0.700 & 0.079 & 0.090 & 0.000 \\ 
		PS Weighting only & 0.000 & -0.761 & 0.761 & 0.083 & 0.080 & 0.000 \\ 
		IDI (no PS adjustment) & 0.000 & 0.036 & 0.036 & 0.109 & 0.107 & 0.933 \\ 
		IDI (PS Matching) & 0.000 & 0.042 & 0.042 & 0.165 & 0.166 & 0.944 \\ 
		IDI (PS Weighting) & 0.000 & -0.007 & 0.007 & 0.108 & 0.107 & 0.946 \\ 
		
		\addlinespace[0.3em]
		\multicolumn{7}{c}{Treatment effect $= -0.5$} \\ \addlinespace[0.3em]
		
		Naive analysis & -0.475 & -1.038 & 0.563 & 0.089 & 0.096 & 0.000 \\ 
		PS Weighting only & -0.475 & -1.105 & 0.631 & 0.091 & 0.091 & 0.000 \\ 
		IDI (no PS adjustment) & -0.475 & -0.448 & 0.027 & 0.109 & 0.108 & 0.936 \\ 
		IDI (PS Matching) & -0.475 & -0.446 & 0.028 & 0.159 & 0.158 & 0.944 \\ 
		IDI (PS Weighting) & -0.475 & -0.469 & 0.006 & 0.107 & 0.108 & 0.952 \\ 
		
		\addlinespace[0.3em]
		\multicolumn{7}{c}{Treatment effect $= -1$} \\ \addlinespace[0.3em]

		Naive analysis & -0.935 & -1.421 & 0.486 & 0.103 & 0.107 & 0.001 \\ 
		PS Weighting only & -0.935 & -1.495 & 0.560 & 0.103 & 0.103 & 0.000 \\ 
		IDI (no PS adjustment) & -0.935 & -0.932 & 0.003 & 0.118 & 0.115 & 0.946 \\ 
		IDI (PS Matching) & -0.935 & -0.932 & 0.003 & 0.162 & 0.161 & 0.946 \\ 
		IDI (PS Weighting) & -0.935 & -0.941 & 0.006 & 0.116 & 0.117 & 0.949 \\ 
		
		\addlinespace[0.3em]
		\hline \hline \addlinespace[0.3em]
		
		\multicolumn{7}{c}{\textbf{Second Case: Covariate-dependent truncation} ($R \mid X$)} \\
		\addlinespace[0.3em] \hline \hline \addlinespace[0.3em]
		
		\multicolumn{7}{c}{Treatment effect $= 0$} \\ \addlinespace[0.3em]
		
		Naive analysis & 0.000 & -0.740 & 0.740 & 0.082 & 0.091 & 0.000 \\ 
		PS Weighting only & 0.000 & -0.687 & 0.687 & 0.092 & 0.090 & 0.000 \\ 
		IDI (no PS adjustment) & 0.000 & -0.056 & 0.056 & 0.117 & 0.116 & 0.915 \\ 
		IDI (PS Matching) & 0.000 & 0.004 & 0.004 & 0.188 & 0.185 & 0.949 \\ 
		IDI (PS Weighting) & 0.000 & -0.010 & 0.010 & 0.134 & 0.132 & 0.948 \\ 
		
		\addlinespace[0.3em]
		\multicolumn{7}{c}{Treatment effect $= -0.5$} \\ \addlinespace[0.3em]
		
		Naive analysis & -0.467 & -1.091 & 0.623 & 0.092 & 0.098 & 0.000 \\ 
		PS Weighting only & -0.467 & -1.056 & 0.589 & 0.100 & 0.100 & 0.000 \\ 
		IDI (no PS adjustment) & -0.467 & -0.546 & 0.078 & 0.111 & 0.109 & 0.889 \\ 
		IDI (PS Matching) & -0.467 & -0.474 & 0.007 & 0.159 & 0.158 & 0.947 \\ 
		IDI (PS Weighting) & -0.467 & -0.490 & 0.023 & 0.120 & 0.120 & 0.944 \\ 
		
		\addlinespace[0.3em]
		\multicolumn{7}{c}{Treatment effect $= -1$} \\ \addlinespace[0.3em]

		Naive analysis & -0.934 & -1.485 & 0.550 & 0.105 & 0.110 & 0.000 \\ 
		PS Weighting only & -0.934 & -1.465 & 0.531 & 0.112 & 0.113 & 0.001 \\ 
		IDI (no PS adjustment) & -0.934 & -1.031 & 0.096 & 0.118 & 0.117 & 0.875 \\ 
		IDI (PS Matching) & -0.934 & -0.960 & 0.026 & 0.159 & 0.158 & 0.946 \\ 
		IDI (PS Weighting) & -0.934 & -0.972 & 0.038 & 0.126 & 0.126 & 0.943 \\ 
		
		\hline
		\multicolumn{7}{l}{\scriptsize $^{*}$Mean estimated log-hazard ratio} \\
		\multicolumn{7}{l}{\scriptsize $^{**}$Absolute bias relative to the true marginal effect} \\
		\multicolumn{7}{l}{\scriptsize $^{***}$Empirical standard deviation} \\
		\multicolumn{7}{l}{\scriptsize $^{\dagger}$Average model-based standard error} \\
		\multicolumn{7}{l}{\scriptsize $^{\ddagger}$Empirical coverage of the 95\% confidence interval} \\
\end{tabular}}
\end{table}

\section{Emulating an Externally Controlled Trial from ECOG-ACRIN 5508} 
\label{sec:application}

We used ECOG-ACRIN 5508 as a case study to illustrate the proposed IDI procedure in an externally controlled survival analysis with delayed treatment initiation. ECOG-ACRIN 5508 was a phase III randomized trial in patients with advanced nonsquamous non-small-cell lung cancer (NSCLC) evaluating maintenance therapy after induction chemotherapy \autocite{ramalingam2019pemetrexed}. The trial is well suited for this illustration because it has a two-stage structure: patients first entered a common induction treatment pathway, and only those who remained eligible after induction were randomized to maintenance therapy.

In the original trial, 1,516 patients were registered and received carboplatin, paclitaxel, and bevacizumab for up to four cycles. Patients without progression after induction were randomized to maintenance therapy with bevacizumab, pemetrexed, or the combination of bevacizumab and pemetrexed. A total of 874 patients, or 57\% of those registered, were randomized to one of the three maintenance groups. Overall survival in the randomized trial was indexed at randomization, which corresponds to the post-induction maintenance decision time. For this illustration, we focused on the comparison between pemetrexed maintenance and bevacizumab maintenance, with bevacizumab serving as the control group.

We used ECOG-ACRIN 5508 to construct an emulated externally controlled trial in which the randomized maintenance comparison provides a benchmark. The purpose of this analysis was not to draw new clinical conclusions about maintenance therapy in NSCLC, but to evaluate whether IDI can reduce the discrepancy between a naive externally controlled analysis and the randomized benchmark when index dates are unobserved in the external-control cohort. The example should therefore be interpreted as a controlled methodological illustration of index-date and eligibility alignment, rather than as an independent comparative effectiveness analysis.

The emulated single-arm trial was defined as the pemetrexed maintenance arm. For these patients, the maintenance decision time was observed and was treated as the post-baseline decision time \(R\). Survival was available both from trial registration and from maintenance initiation. Membership in the pemetrexed maintenance arm required that patients survive and remain progression-free through induction, so the observed single-arm cohort represents a post-initiation population selected by both survival and maintenance eligibility.

The external-control source cohort was constructed from patients assigned to bevacizumab maintenance and patients who were registered to the trial but did not proceed to maintenance randomization. Including patients who did not reach maintenance randomization was intentional. IDI is designed for external-control sources indexed at an earlier clinical milestone, where the source population may include patients who would not survive or remain eligible to the later treatment-decision time. In this emulation, the analogous maintenance decision time was treated as unobserved for all external-control patients, and survival was indexed at trial registration. This construction creates the index-date and eligibility-alignment problem that IDI is designed to address.

To emulate covariate imbalance typical of externally controlled analyses, we sampled with replacement from the external-control source cohort to induce differences in baseline prognostic factors relative to the pemetrexed single-arm cohort. The resampling was based on disease stage, ECOG performance status, gender, smoking history, and bone metastasis status. The resulting baseline covariate distributions in the emulated single-arm and external-control cohorts are shown in Table~\ref{tab:tab2-5508}. 

In ECOG-ACRIN 5508, eligibility for maintenance randomization required absence of progression after induction. Therefore, this application used the eligibility-extended version of IDI described in Section~\ref{sec:methods}. Specifically, external-control patients were required to remain alive and progression-free at the imputed maintenance decision time \(R^*\), reflecting the trial requirement that patients be eligible for maintenance randomization after induction therapy. Thus, the truncation and eligibility condition was based on
\[
E(R^*)=\mathbb{I}\{T>R^*,P>R^*\},
\]
where \(P\) denotes time to progression from trial registration. Progression information was available for randomized bevacizumab patients. For patients who did not proceed to maintenance randomization, their failure to enter the randomized maintenance phase was used to encode that they did not satisfy the maintenance-eligibility process. Implementation details are provided in Section~D of the Supporting Information.

The resulting emulated single-arm and external-control cohorts included 279 and 695 patients, respectively. We compared two approaches. The naive analysis compared the pemetrexed single-arm cohort with the emulated external-control cohort using survival indexed at trial registration, without covariate adjustment, index-date alignment, or eligibility restriction at the maintenance decision time. The IDI analysis used the propensity score weighting implementation described in Section~\ref{sec:methods}. Specifically, we estimated the maintenance-decision-time distribution from the pemetrexed arm, imputed a pseudo-maintenance decision time for each external-control patient, excluded external-control patients whose death, progression, or end of follow-up occurred before the imputed decision time, and applied ATT weighting to adjust for measured differences in age, gender, disease stage, ECOG performance status, smoking history, and bone metastasis status. We used 1,000 nonparametric bootstrap replicates to account for uncertainty from index-date imputation and nuisance-parameter estimation.

We summarized the comparison using both a weighted Cox model and marginal survival contrasts. The Cox model provides a familiar model-based summary of the aligned survival curves, but it is not the primary estimand-defining quantity. Because IDI targets marginal post-initiation survival functions, we also estimated survival probability differences at clinically interpretable landmark times. Specifically, for a landmark time \(\tau\), we estimated
\[
\Delta_S(\tau) = \hat S_1(\tau)-\hat S_0(\tau),
\]
where \(\hat S_1(\tau)\) is the estimated pemetrexed survival probability from the observed maintenance decision time and \(\hat S_0(\tau)\) is the estimated control survival probability from the imputed decision time after IDI, eligibility restriction, and ATT weighting. Confidence intervals for \(\Delta_S(\tau)\) were obtained using the same nonparametric bootstrap procedure. In the results below, we report survival probability differences at \(\tau= 12\) months and \(\tau=36\) months, together with the hazard ratio as a secondary summary.

\begin{figure}[t]
	\centering
	\includegraphics[width=0.8\linewidth]{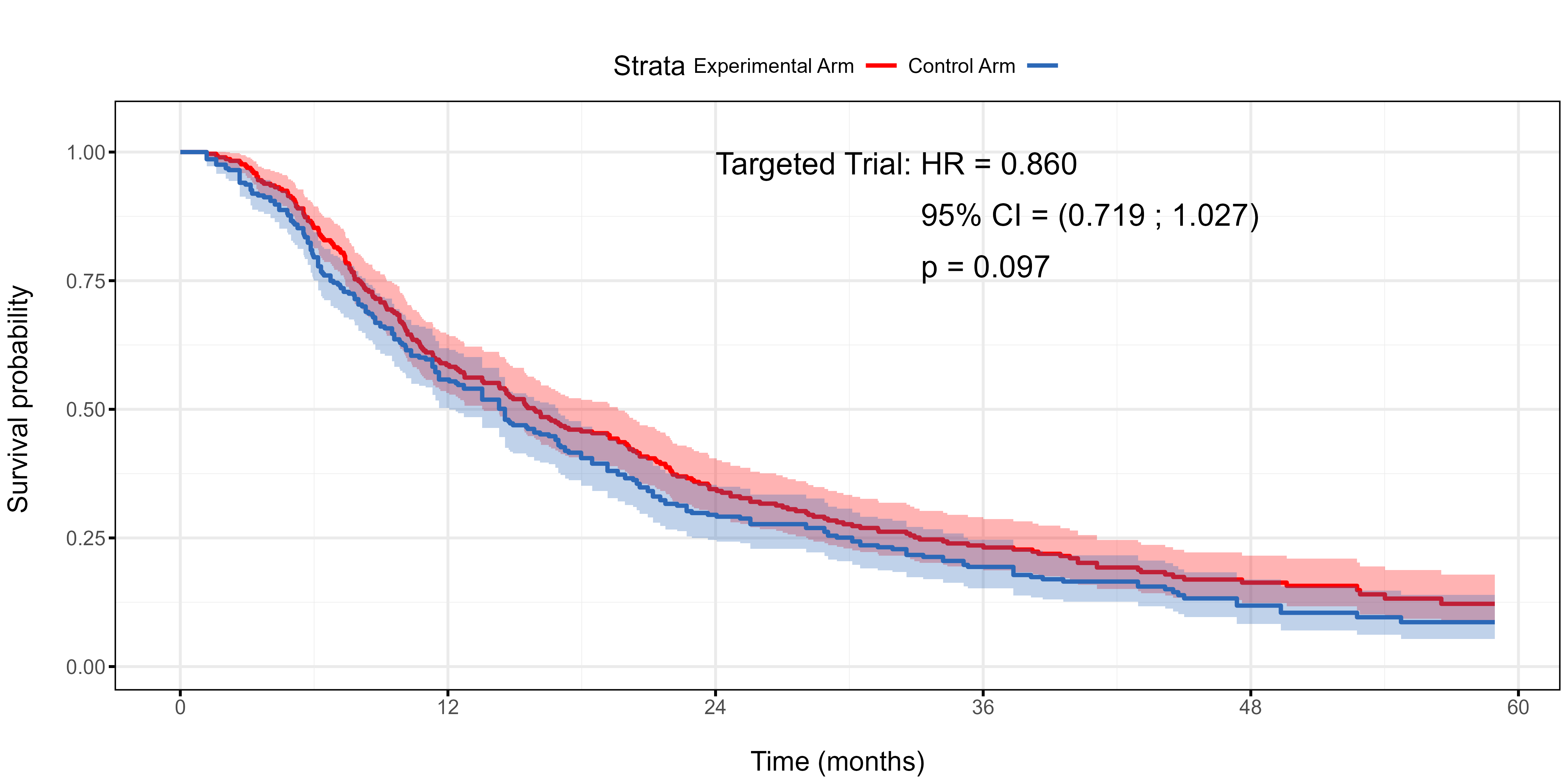}
	\includegraphics[width=0.8\linewidth]{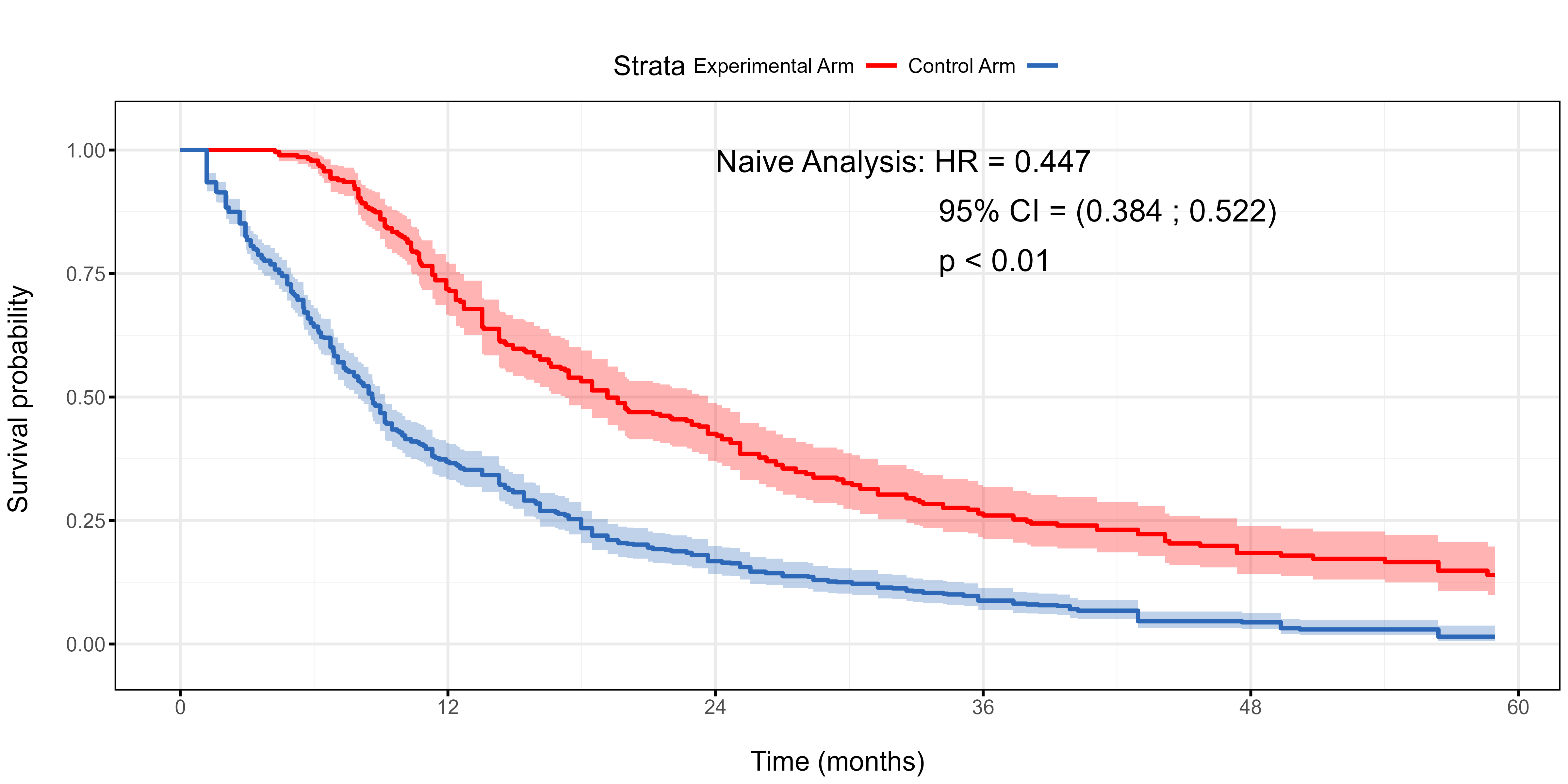}
	\includegraphics[width=0.8\linewidth]{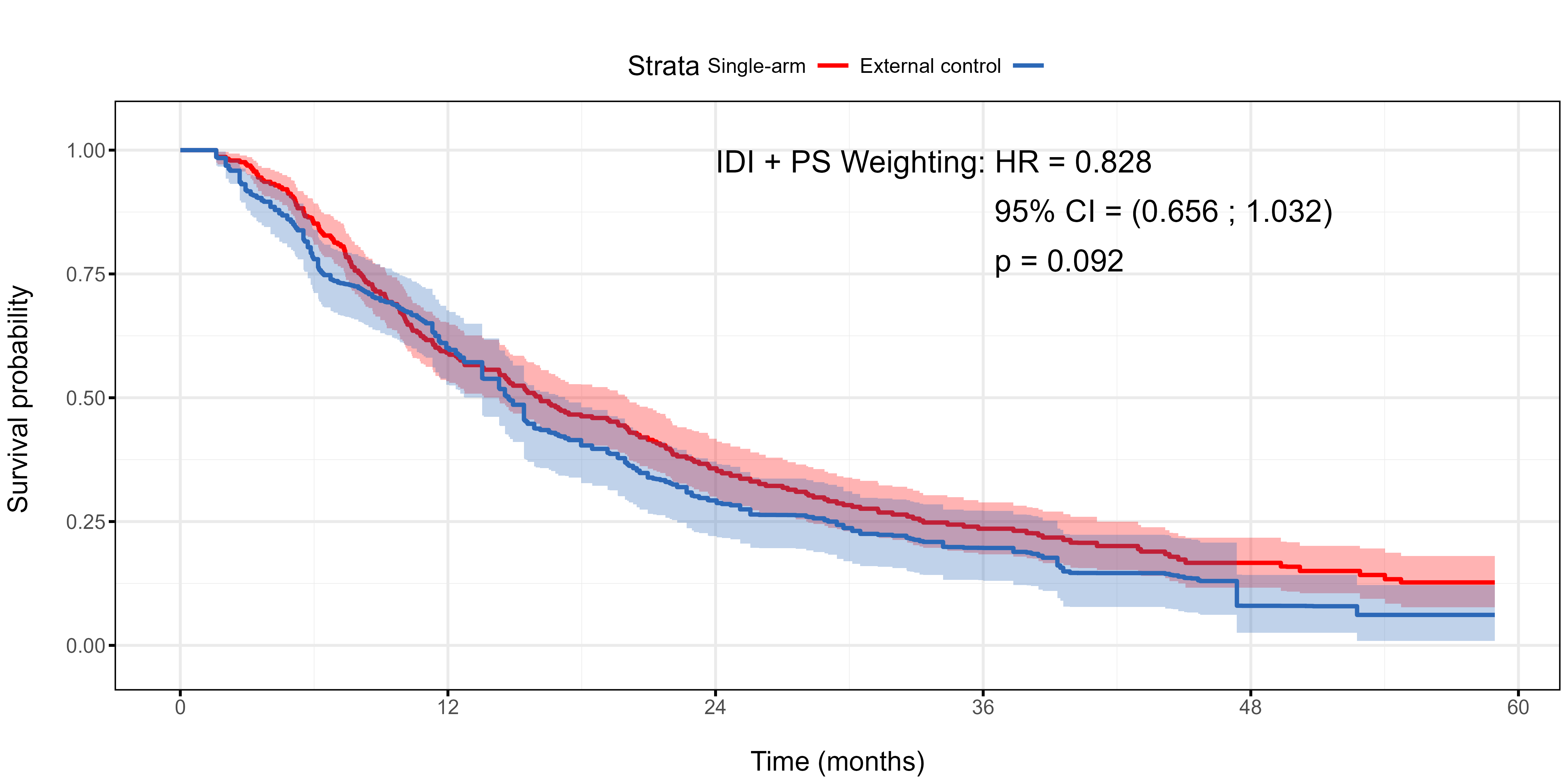}
	\caption{ECOG-ACRIN 5508 emulated ECT analysis. Top: randomized benchmark comparing pemetrexed with bevacizumab maintenance therapy, indexed at maintenance randomization. Middle: naive emulated ECT comparison indexed at trial registration. Bottom: IDI analysis using propensity score weighting after imputed index-date alignment, risk-set and eligibility restriction, and covariate adjustment.}
	\label{fig:res_ecog_idi}
\end{figure}

Results from the randomized benchmark, naive emulated ECT analysis, and IDI analysis are shown in Figure~\ref{fig:res_ecog_idi} and summarized in Table~\ref{tab:ecog_results}. In our analytic dataset and using the same Cox modeling convention as in the emulated analyses, the randomized benchmark hazard ratio for pemetrexed versus bevacizumab was 0.860 (95\% CI: 0.719--1.027; \(p=0.097\)). The corresponding landmark survival probability differences were 0.032 (95\% CI: -0.049--0.113) at 12 months and 0.042 (95\% CI: -0.026--0.109) at 36 months. The naive externally controlled analysis strongly favored the pemetrexed single-arm cohort, with an estimated hazard ratio of 0.447 (95\% CI: 0.384--0.522; \(p<0.01\)). The naive landmark survival probability differences were 0.343 (95\% CI: 0.278--0.408) at 12 months and 0.170 (95\% CI: 0.113--0.226) at 36 months. This discrepancy is consistent with index-date misalignment, lack of analogous maintenance-eligibility restriction in the external-control cohort, and induced covariate imbalance.

After applying IDI with propensity score weighting, the estimated hazard ratio was 0.828 (95\% CI: 0.656--1.032; \(p=0.092\)), which was closer to the randomized benchmark. The IDI-based landmark survival probability differences were -0.010 (95\% CI: -0.101--0.086) at 12 months and 0.039 (95\% CI: -0.050--0.127) at 36 months, also moving toward the randomized benchmark relative to the naive analysis. These results suggest that, in this controlled emulation, IDI reduced the discrepancy introduced by using an external-control source indexed at an earlier clinical milestone.

This application illustrates how IDI can be used to reconstruct a post-initiation comparison from data indexed at an earlier clinical milestone. In this emulated ECT, IDI substantially reduced the discrepancy between the naive estimate and the randomized benchmark. The example also highlights an important practical requirement: when eligibility at treatment initiation depends on clinical events beyond survival, such as progression status, those events must be available or credibly approximated in the external-control data to emulate the post-initiation target comparison. When such information is unavailable or non-uniformly measured, IDI can still align the index date, but the resulting analysis should be interpreted as an approximation whose validity depends on the quality of the eligibility approximation.

\begin{table}[t]
	\centering
	\singlespacing
	\caption{Baseline covariate distribution in the emulated ECOG-ACRIN 5508 single-arm and external-control cohorts before IDI adjustment. Values are mean (SD) for age and \(n\) (\%) for categorical variables. Percentages may not sum to 100 because of rounding.}
	\label{tab:tab2-5508}
	\small
	\renewcommand{\arraystretch}{1.08}
	\begin{tabular}{lcc}
		\toprule
		Variable & Single-arm (n = 279)  &  External control (n = 695)  \\ 
		\midrule
		Age, mean (SD)  & 63.0 (9.2)    & 65.9 (8.3) \\ 
		\addlinespace[0.1em]
		
		Gender &  &  \\ 
		\hspace{1cm}Male & 137 (49.1)  & 373 (53.7) \\ 
		\hspace{1cm}Female & 142 (50.9)  & 322 (46.3) \\  
		\addlinespace[0.1em]
		
		Stage   & & \\ 
		\hspace{1cm}IIIB & 4 (1.5) & 12 (1.7) \\ 
		\hspace{1cm}IV M1a & 88 (31.5) & 70 (10.1) \\ 
		\hspace{1cm}IV M1b & 170 (60.9)  & 601 (86.5) \\ 
		\hspace{1cm}Recurrent & 17 (6.1)  & 12 (1.7) \\ 
		\addlinespace[0.1em]
		
		ECOG performance status &  &  \\ 
		\hspace{1cm}0 & 131 (47.0)  & 152 (21.9) \\ 
		\hspace{1cm}1 & 148 (53.0)  & 543 (78.1) \\  
		\addlinespace[0.1em]
		
		Smoking history   & & \\ 
		\hspace{1cm}Current & 132 (47.3)& 378 (54.4) \\ 
		\hspace{1cm}Former& 116 (41.6) & 265 (38.1) \\ 
		\hspace{1cm}Never & 31 (11.1)  & 52 (7.5) \\ 
		\addlinespace[0.1em]
		
		Bone metastasis   & & \\ 
		\hspace{1cm}Yes & 178 (63.8)& 250 (36.0) \\ 
		\hspace{1cm}No & 101 (36.2) & 445 (64.0) \\ 
		\bottomrule
	\end{tabular}
\end{table}

\begin{table}[t]
	\centering
	\caption{ECOG-ACRIN 5508 emulated ECT results compared with the randomized benchmark}
	\label{tab:ecog_results}
	\small
	\renewcommand{\arraystretch}{1.12}
	\begin{tabular}{lccc}
		\toprule
		Analysis & Hazard ratio (95\% CI) & \(\Delta_S(\tau_1)\) (95\% CI) & \(\Delta_S(\tau_2)\) (95\% CI) \\
		\midrule
		Randomized benchmark & 0.860 (0.719--1.027) & 0.032 (-0.049--0.113) & 0.042 (-0.026--0.109) \\
		Naive emulated ECT & 0.447 (0.384--0.522) & 0.343 (0.278--0.408) & 0.170 (0.113--0.226) \\
		IDI (PS weighting) & 0.828 (0.656--1.032) & -0.010 (-0.101--0.086) & 0.039 (-0.050--0.127) \\
		\bottomrule
	\end{tabular}
	\begin{flushleft}
		\footnotesize
		\(\Delta_S(\tau)=\hat S_1(\tau)-\hat S_0(\tau)\), where positive values favor pemetrexed maintenance. Landmark times are \(\tau_1 = 12\) months and \(\tau_2 = 36\) months from the observed or imputed maintenance decision time.
	\end{flushleft}
\end{table}

\section{Discussion} 
\label{sec:discussion}

This work addresses index-date misalignment in externally controlled single-arm trials with delayed treatment initiation. In such settings, treatment initiation times are unobserved in the external-control cohort, while observed initiation times in the treated cohort are themselves subject to truncation because initiation is observed only among patients who survive and remain eligible long enough to initiate treatment. IDI addresses this asymmetry by estimating the marginal initiation-time distribution in the target single-arm population and using it to impute comparable index dates for external controls. The approach is designed for post-initiation survival comparisons in which follow-up begins at a later treatment-decision time, rather than at an earlier clinical milestone.

A distinctive feature of IDI is that the estimated initiation-time distribution plays two related roles. It is used to impute index dates for external controls, and it is also used to account for selection induced by delayed initiation when estimating population-membership propensity scores. This coupling reflects the observed data structure: the single-arm cohort represents patients who have already survived and remained eligible to the later decision time. Therefore, directly using the empirical distribution of observed initiation times or directly fitting propensity scores in the observed treated cohort can target a survivor-selected population rather than the intended pathway-entry target population.

In many externally controlled single-arm settings, the average treatment effect in the treated population is a natural target because the single-arm trial defines the population for which the investigational treatment is being evaluated. Accordingly, the proposed implementation focuses on ATT estimation. Other marginal estimands, including the average treatment effect, the average treatment effect in the untreated, or the average treatment effect in the overlap population, can in principle be obtained by modifying the weighting scheme. These estimands correspond to different target populations and should be used only when they align with the scientific question and the available data. In particular, overlap weighting may improve numerical stability when covariate overlap is limited, but it targets an overlap population rather than the single-arm trial population \autocite{li2019addressing}.

The nonparametric estimator of the marginal initiation-time distribution used in this paper relies on a working independence assumption between \(R\) and baseline covariates \(X\). This assumption is not part of the definition of the post-initiation survival contrast. Rather, it is an estimation assumption used to recover \(F_R\) from truncated single-arm data. When initiation timing depends strongly on baseline covariates, a conditional model for \(F_R(\cdot\mid X)\) that explicitly accounts for truncation would be preferable. The simulation results suggest that the proposed estimator can be reasonably robust in the scenarios considered, but they do not eliminate the need for conditional initiation-time models in settings where treatment timing is strongly driven by prognostic characteristics. Developing flexible estimators of \(F_R(\cdot\mid X)\) is an important direction for future work.

Confounding adjustment in the proposed framework was implemented using propensity score weighting or matching, reflecting common practice in externally controlled comparisons and providing a transparent link between covariate balance and the target estimand. The framework is not tied to propensity scores. Alternative balancing approaches, such as entropy balancing, may also be used when they target the same population and adequately address measured differences between cohorts \autocite{hainmueller2012entropy}. Regardless of the adjustment method, the truncated nature of the observed single-arm cohort should be considered when constructing weights or balance constraints. Covariate balance and overlap diagnostics, including standardized mean differences and prognostic-score summaries, remain essential, although they cannot establish exchangeability.

Index-date alignment should be distinguished from eligibility alignment. IDI is designed to align treatment assignment time and time zero in settings with delayed initiation. It does not replace longitudinal eligibility information. Rather, it defines the imputed decision time at which such information should be evaluated in the external-control cohort. In the notation used above, external-control patients should satisfy the relevant condition \(E(R^*)=1\) at the imputed decision time. In some settings, this condition may require only survival to \(R^*\). In others, it may also require absence of progression, adequate organ function, recovery from induction therapy, or other time-varying clinical criteria. When this information is sparse, nonuniformly collected, or unavailable, IDI can still address time-zero alignment, but the resulting analysis should be interpreted as an approximation to the intended post-initiation comparison.

The proposed approach is closely related to recent work emphasizing index-date alignment in external-control and precision-medicine studies. Methods such as prescription-time distribution matching and related index-date emulation approaches assign pseudo-index dates to controls by resampling observed initiation times in treated patients\autocite{zhou2005survival,antunes2026choosing,rippin2025external}. These approaches may perform well when truncation is limited. However, when initiation times are observed only among patients who survive and remain eligible to initiate treatment, the empirical distribution of observed initiation times estimates a truncated, survivor-selected distribution. The focused simulation comparing IDI with PTDM-type imputation showed that the two approaches behaved similarly under mild truncation, but diverged as truncation became more severe. Under heavier truncation, PTDM-type imputation assigned systematically earlier pseudo-index dates to external controls and produced increasing bias, whereas IDI remained approximately unbiased by estimating the marginal initiation-time distribution. Detailed results are provided in Section~C of the Supporting Information.

The IDI framework targets marginal survival functions indexed at the observed or imputed treatment-decision time. Therefore, the survival summary used to report treatment effects should be chosen to reflect the scientific question. Hazard ratios remain familiar and were used in the simulations and application as model-based summaries, but they are non-collapsible and may be difficult to interpret under nonproportional hazards. Marginal summaries such as restricted mean survival time differences or landmark survival probability differences provide more direct contrasts of the survival functions targeted by IDI. For this reason, the ECOG-ACRIN 5508 illustration reports both hazard ratios and landmark survival probability differences.

Several limitations should be noted. First, patient-level index-date imputation introduces stochastic variability. We used a nonparametric bootstrap to account for uncertainty from index-date imputation and nuisance-parameter estimation. Alternative approaches, such as multiple imputation combined using Rubin's rules, could also be considered \autocite{rubin2018multiple}. Second, enforcing the risk-set and eligibility condition at the imputed index date may exclude external-control patients whose event, progression, or follow-up end occurs before \(R^*\). This can reduce the effective sample size and diminish power, especially when initiation is delayed or external-control follow-up is short. Third, if censoring occurs before the imputed index date, the observed retention condition becomes \(Y>R^*\) rather than \(T>R^*\), which can bias estimation and induce post-truncation imbalance. Addressing this issue may require censoring-aware estimators or alternative approaches that use baseline survival information without discarding patients censored before the imputed decision time.

We also assumed that assignment to the later-phase treatment strategy coincides with receipt of that strategy. This assumption is natural in many single-arm trial settings, where enrolled patients initiate the investigational regimen by design. It may not hold when non-initiation, nonadherence, treatment switching, or early discontinuation is common. In such settings, the target estimand should distinguish assignment effects, treatment-receipt effects, and per-protocol effects before IDI is applied. The feasibility of these estimands depends on whether the single-arm and external-control data contain sufficient treatment-process information.

As in other externally controlled studies, the proposed approach depends on the quality and comparability of measured data. Baseline covariates must be measured with sufficient completeness to support adjustment, and outcome ascertainment must be comparable across cohorts. When baseline covariates are missing, multiple imputation could in principle be combined with IDI, although recent work suggests that some missing-data strategies can perform poorly when missingness is concentrated in the external-control cohort \autocite{rippin2025external}. Missing or inconsistently measured longitudinal eligibility information is especially consequential because eligibility must be evaluated at the imputed decision time, not only at pathway entry.

Although the motivating examples arise in oncology and rare diseases, the method is not disease-specific. IDI may be useful in time-to-event settings where one cohort is indexed at treatment initiation, the comparator source is indexed at an earlier milestone, and observed initiation times are truncated by survival or eligibility to initiation. Examples include multi-phase treatment regimens, maintenance therapy, transplantation studies, and eligibility-driven adaptive treatment settings. Throughout this work, we use the term ``externally controlled trials'' to describe studies comparing a single-arm trial with an external-control cohort, while acknowledging recent distinctions between externally controlled trials and external comparator cohort studies \autocite{rippin2024external}. The methodological considerations discussed here apply broadly to externally sourced control comparisons when delayed initiation and index-date misalignment are present.

Future work should focus on practical extensions needed for broader use of IDI. These include conditional estimation of \(F_R(\cdot\mid X)\) when initiation timing depends on patient characteristics, methods for handling censoring before the imputed index date, and sensitivity analyses for incomplete or nonuniform eligibility information at \(R^*\). These developments would extend IDI beyond the core setting studied here while preserving its main purpose: aligning the post-initiation decision time before comparing survival outcomes.

\section*{Acknowledgments}
This manuscript was prepared using data from Datasets NCT01107626 from the NCTN/NCORP Data Archive of the National Cancer Institute’s (NCI’s) National Clinical Trials Network (NCTN). Data were originally collected from clinical trial NCT number NCT01107626 "Bevacizumab or Pemetrexed Disodium Alone or In Combination After Induction Therapy in Treating Patients With Advanced Non-Squamous Non-Small Cell Lung Cancer". All analyses and conclusions in this manuscript are the sole responsibility of the authors and do not necessarily reflect the opinions or views of the clinical trial investigators, the NCTN, the NCORP or the NCI.

\printbibliography


\appendix

\label{lastpage}

\clearpage

\includepdf[pages=-]{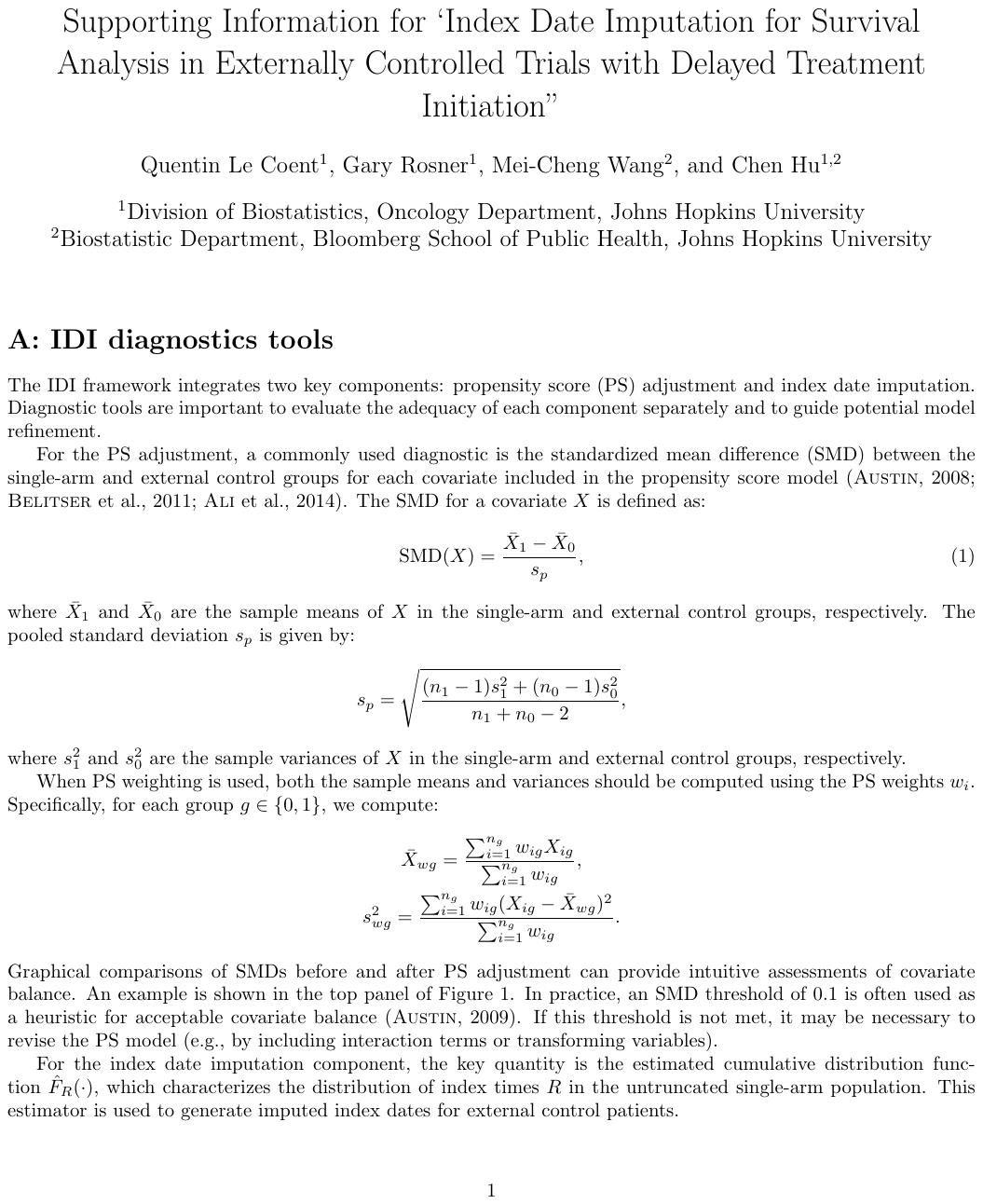}

\end{document}